\documentclass[12pt,preprint]{aastex}
\usepackage{psfig} 

\slugcomment{To appear in ApJ Letters Vol. 570, 10 May 2002.}

\shorttitle{Weak lensing galaxy by galaxy}
\shortauthors{A.\,W. Blain}

\begin{document}


\title{Detecting gravitational lensing cosmic shear from samples of several 
galaxies using two-dimensional spectral imaging} 


\author{A.\,W. Blain}
\affil{Astronomy Department, Caltech 105-24,
    Pasadena, CA 91125}
\email{awb@astro.caltech.edu} 



\begin{abstract}
Studies of weak gravitational lensing by 
large-scale structures require the measurement of
the distortions introduced to the shapes of distant galaxies at the few percent 
level by anisotropic light deflection 
along the line of sight. In order to detect this signal on 1'--10'
scales in a particular field, 
accurate measurements of correlations between the  
shapes of order $10^3$--$10^4$ galaxies are required. 
This large-scale averaging is required to accommodate 
the unknown intrinsic shapes of the background galaxies, even with careful 
removal of systematic effects. 
Here an alternative is discussed. If it is possible to measure 
accurately the
detailed dynamical structure of the background galaxies, in
particular rotating disks, then it should be possible to
measure directly the cosmic shear distortion since it generally 
leads to a non-self-consistent
rotation curve. Narrow spectral lines and an excellent 
two-dimensional spatial resolution
are required.
The ideal lines are CO rotational transitions, and the ideal telescope 
is the Atacama Large Millimeter Array (ALMA).
\end{abstract}




\section{Introduction}

Weak gravitational lensing is a valuable tool for studying the
mass distribution along and around the line of sight to distant galaxies 
(Bartelmann \& Schneider 2001). Away from galaxies  
and clusters, the
effect is small: the axis ratio of high-redshift 
galaxies is modified by a few percent. Because
the intrinsic shapes of distant galaxies
are not known, it is necessary to stack many thousands of galaxy images  
to detect a signal. This requires large, high-quality
images, and so cosmic shear can only be detected statistically over 
areas extending more than about 0'.5. 
Cosmic shear
has been detected in both deep optical images   
(Bacon, Refregier \& Ellis 2000; Wittman et al.\ 2000; Maoli et al.\ 2001; 
Pirzkal et al.\ 2001; 
van Waerbeke et al.\ 2001; Wilson, Kaiser \& Luppino 2001) 
and very wide, shallow optical surveys 
(Fischer et al.\ 2000), but  
at present it is not possible
to measure the shear towards a specific point
on the sky. 

Here we investigate the possibility
of detecting weak lensing in observations of {\em individual} 
galaxies, by obtaining  
high-fidelity, two-dimensional, resolved spectroscopic measurements of 
morphology and dynamics. This may
be possible using an integral field unit (IFU) adaptive optics (AO) 
optical or 
near-infrared(IR) spectrograph to image the narrow stellar absorption and 
hydrogen recombination lines. 
It should definitely be possible using the 
interferometric imaging of CO  
rotation line 
emission at millimeter 
wavelengths using the 
Atacama Large Millimeter
Array (ALMA; Wootten 2001), or using meter-wave H{\sc i} imaging using  
a Square Kilometer Array 
(SKA)\footnote{www.ras.ucalgary.ca/SKA/ska\_science.shtml}
radio interferometer. 

Measuring cosmic shear galaxy by galaxy would provide
many new scientific opportunities, and a very valuable check on
systematic errors in conventional techniques.
Along several closely spaced lines of sight, a 
tomographic survey of the mass distribution out  
into the distant Universe could be made. The properties of the weak lensing 
shear field on very small (arcsec) galactic scales are unknown. 
Substructure in 
dark matter halos could lead to significant variations in the  
shear field on these scales. This would be washed out  
by averaging over both 
arcminute-scale fields in existing cosmic shear surveys  
(Pirzkal et al.\ 
2001) and the stacked images analyzed for galaxy-galaxy 
lensing around the positions of bright galaxies (McKay et al.\, 2002). 

\section{The effect of weak lensing shear on rotation curves}

\begin{figure}
\psfig{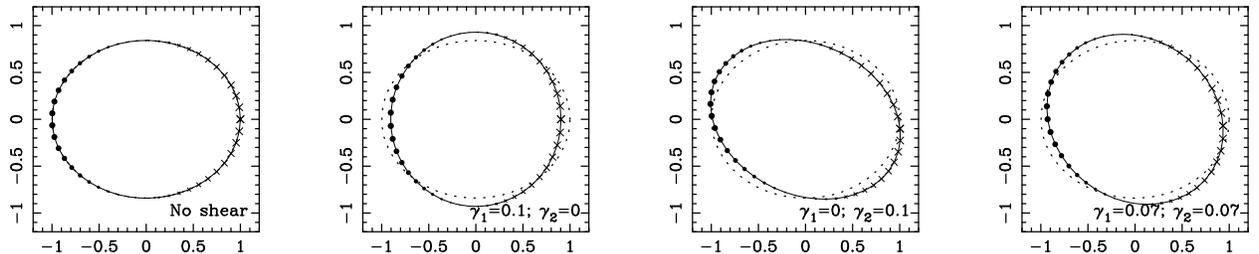}
\caption{Position and line-of-sight velocity of points ({\it filled circles}, 
approaching; {\it crosses}, receding) on 
a rotating ring with $i=0.57$\,rad/$0^\circ.33$ and $\rm{P.A.}=0$\,rad, 
with no weak lensing ({\it leftmost panel}) and for three 
large values of weak-lensing shear 
$\vert \vec \gamma \vert
= 0.1$ 
({\it three rightmost panels}). The unlensed 
ring is shown by the dotted line for comparison. 
\label{fig1}}
\end{figure}

\begin{figure}
\psfig{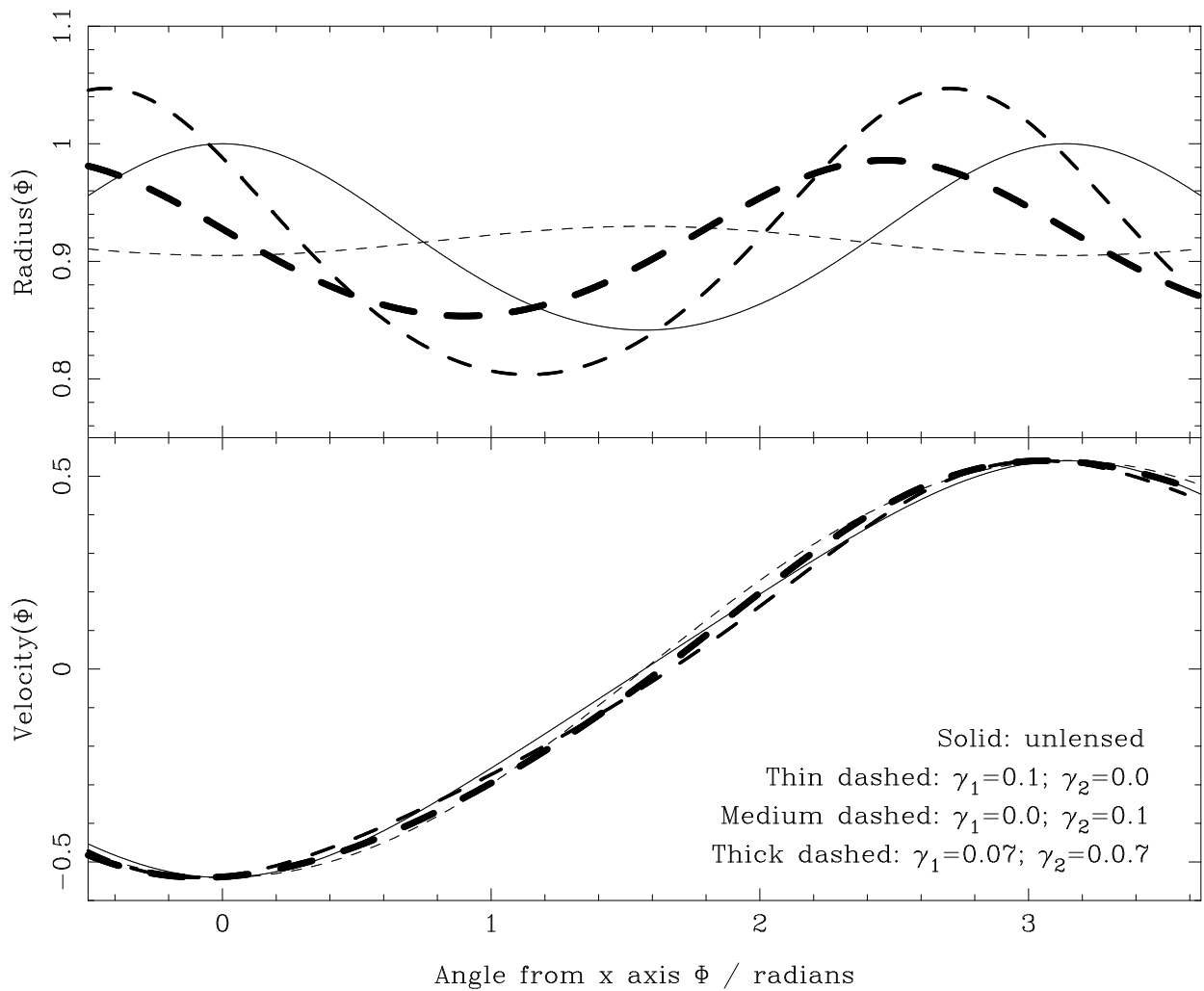}
\caption{Radial distance and line-of-sight 
velocity of points on the ring shown in Fig.\,1 as a function of angle 
on the sky $\Phi$ over more than half of a circle. 
The solid and dashed lines show the results with zero and nonzero shear, 
respectively. 
Lensing has a more 
significant effect on radial position than 
velocity. 
\label{fig2}}
\end{figure}

We consider the illustrative case of representing a 
galaxy as an inclined rotating  
circular ring centered on the origin, with an inclination angle  
$i$ between the rotation axis and the line of sight. The 
position angle (P.A.) of the projected elliptical ring on the sky is  
free to vary. The line-of-sight velocity of a point on the ring 
$v \propto  
{\rm sin}\,i \, {\rm cos}\,\Phi$, where $\Phi$ is the angle 
between the point and 
the major axis of the ellipse. 
The dependence of the 
radial position and the line-of-sight 
velocity
of a point on the ring on $\Phi$ are shown for a model with 
$i = 0.57$\,rad\,($\simeq 33^\circ$) 
in Figures\,\ref{fig1} and \ref{fig2}. 

Weak gravitational lensing shears the ellipse without
changing its area (Schneider, Ehlers \& Falco 1992).
The transformation of angular position  
$(\theta_{\rm x},
\theta_{\rm y})$ to
$(\theta'_{\rm x}, \theta'_{\rm  y})$ due to lensing by a shear field 
$\vec{\gamma}=(\gamma_1,
\gamma_2)$ is described by
\begin{eqnarray}
\left(
\begin{array}{c}
\theta'_{\rm x} \\
\theta'_{\rm y} 
\end{array} \right) = 
\left( 
\begin{array}{cc}
1-\kappa-\gamma_1 & - \gamma_2 \\
- \gamma_2 & 1-\kappa+\gamma_1
\end{array}	\right)
\left(
\begin{array}{c}
\theta_{\rm x} \\
\theta_{\rm y} 
\end{array} \right). 
\end{eqnarray}
In the weak lensing regime, $\vec{\gamma}$ is small and the 
convergence $\kappa \simeq 1 - \sqrt{1 + \vert \gamma \vert^2} \simeq 0$, with 
$\vert \gamma \vert^2 = \gamma_1^2 + \gamma_2^2$. 
The $\gamma_2=0$ shear component is 
aligned along $\theta_{\rm x} = 0$, while the $\gamma_1=0$ component 
is aligned along $\theta_{\rm x}=\theta_{\rm y}$. 
The frequency of radiation is not affected by lensing, and so 
the line-of-sight velocity remains the same. 
The velocity and position of a ring is thus modified
by weak lensing in a straightforward way.\footnote{The velocity field of 
strongly lensed giant arcs in clusters, for which $\kappa$ 
is large, was discussed by 
Narasimha \& Chitre (1993).} 

Weak lensing by a large 10\% shear field
($\vert \gamma \vert = 0.1$) 
aligned in different directions is illustrated in  
Figures\,\ref{fig1} and \ref{fig2} 
for a galaxy with 
${\rm P.A.}=0$ and $i=0.57$\,rad. In most orientations, 
the effect of lensing is similar
to modifying the P.A. and $i$. 
If $\vec{\gamma}$ is perpendicular to the projected 
rotation axis, as shown in the second to leftmost panel 
of Figure\,\ref{fig1}, then 
the effect of shear is indistinguishable from  
modifying $i$. Based on position data alone, there is always a 
degeneracy between $\vec{\gamma}$, $i$ and P.A. However, 
velocity data reduces the degeneracy significantly. 
As shown by the bottom panel of Figure\,\ref{fig2}, the velocity field 
defines P.A. accurately, even when $\vec{\gamma}$ is large.\footnote{This 
of course 
requires that the measured velocity field traces the general potential 
of the galaxy and is not affected strongly by a bar, a warp or a `lop-sided' 
$m_1$ asymmetry.} 
Shear affects the position much more 
significantly than the velocity. If $\vec{\gamma}$ 
is not 
perpendicular to the rotation axis, then the sheared ellipse cannot 
be described self-consistently by a rotating ring: for example, neither  
shifting the 
radial curves in Figure\,\ref{fig2} to the left or right, 
as expected for changing 
P.A., nor modifying the amplitude of their modulation, as
expected for changing $i$, can fully mimic the effects of lensing 
if $\gamma_2 \ne 0$. 

\begin{figure} 
\psfig{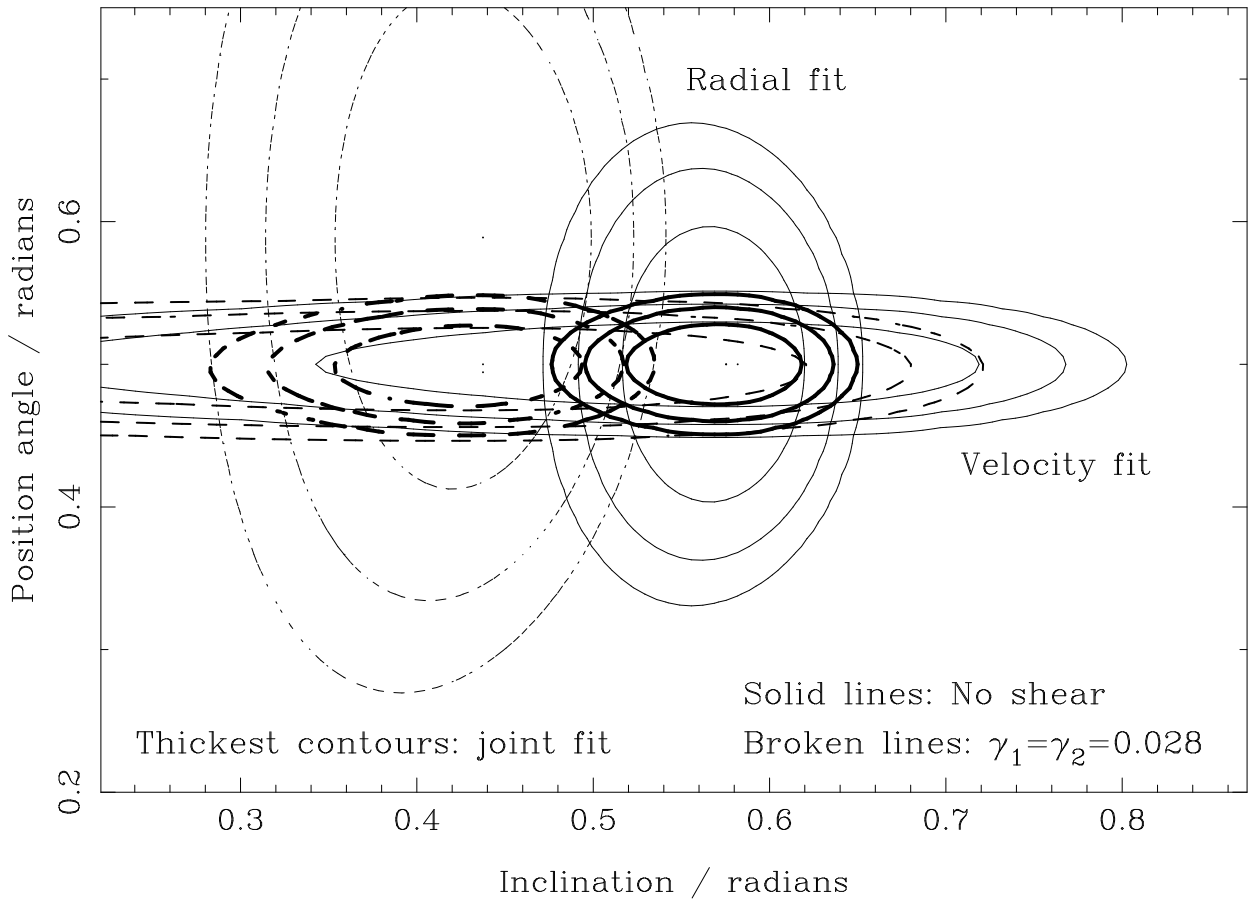}
\caption{Likelihood ($P$) contours obtaining by fitting 
the P.A. and $i$ to simulated data from a rotating ring with P.A.$=0.5$ and 
$i = 0.57$\,rad, 
based on 100 data points with 5\% 
accuracy for both position and velocity. The contours are all  
spaced by factors of 10 away from the peak value, and so the  
second contour corresponds 
approximately to a 3$\sigma$ error. With non-zero shear 
the quality of the joint radius-velocity fit to 
the P.A. and $i$ is reduced from 
$P \simeq 2.4\times 10^{-5}$ to 
$1.2 \times 10^{-5}$. 
The velocity data always fixes 
the P.A. accurately. 
\label{fig3}}
\end{figure} 

\begin{figure}
\psfig{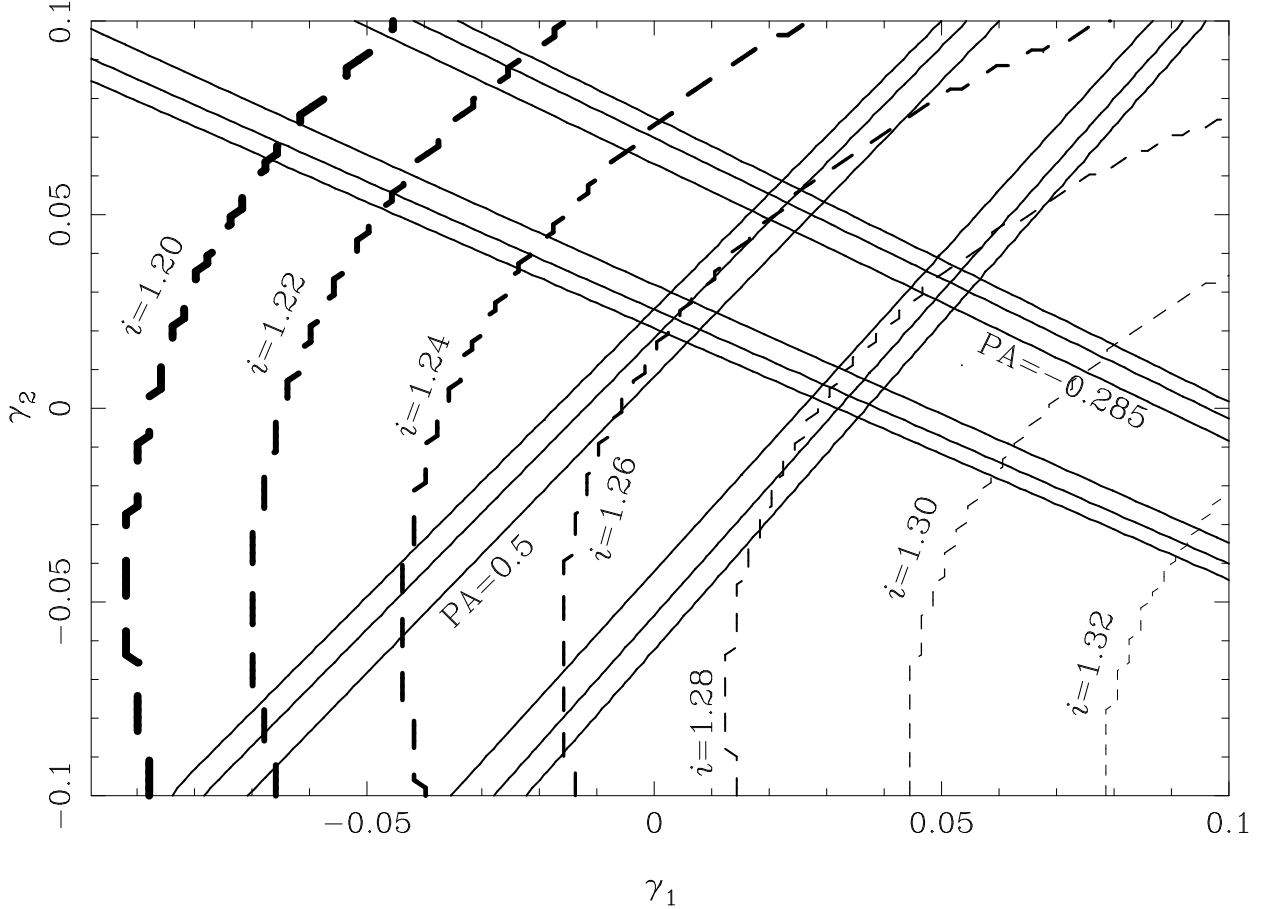}
\caption{Results of fitting a common shear field using simulated 
data for two ring galaxies 
at inclination 
$i=1.27$\,rad (72$^\circ$), with ${\rm P.A.}=0.5$ and $-0.285$\,rad  
($\pi/4$\,rad apart). The uncertainties  
in Fig.\,\ref{fig3} are assumed. The 
solid lines show equal 
probabilities in the $\vec{\gamma}$ plane, and 
are spaced by factors of 10 away from the 
best-fitting values ($P = 7 \times 10^{-5}$). 
The values of $i$ for which the 
joint probability of a fit to $i$ and $\vec{\gamma}$ is maximized are shown 
by the dashed lines for the ${\rm P.A.}=-0.285$ galaxy, 
demonstrating a powerful degeneracy between $i$ and $\vec{\gamma}$. 
Prior information about the maximum value of 
$\vert \vec{\gamma} \vert$ could be used to limit the extent of the 
probability contours in the degenerate direction. 
An unambiguous measurement of the input shear $\vec \gamma = (0.028, 0.028)$ 
is obtained where the solid lines cross. The 
P.A. is determined accurately by the velocity data 
(see Fig.\,\ref{fig3}) and leads 
to very little degeneracy with $\vec{\gamma}$; therefore, it is set  
to its fixed input value in the fitting process. 
\label{fig4}}
\end{figure} 

If sufficiently accurate spatially resolved two-dimensional spectroscopic 
images of a distant rotating galaxy were available,  
then a shear signal should 
be detectable unambiguously without stacking or 
averaging images.
The results of a maximum likelihood fit to the P.A. and $i$, 
taking into account both simulated position and velocity data 
are shown in Figure\,\ref{fig3}, for both zero shear and a  
typical shear value of 4\%.  
An accurate self-consistent fit is obtained for zero shear, whose 
quality is 
significantly improved by including the velocity data, 
especially for low-inclination 
disks ($i \simeq 0$). With shear, 
there is a clear offset between the best-fit 
values of the P.A. 
and $i$ determined from the velocity and position data separately, 
the detection of which could be used to diagnose that 
significant shear is present in real data. The 
probability of the joint velocity-position fit is also reduced by a
factor of about 2. 
For an almost edge-on disk ($i \simeq \pi/2$), the extra information provided  
by velocity measurements is less significant, both for   
determining P.A. and for detecting a shear offset in 
the best-fit parameters. 
Projected circular rotating 
galaxies are good 
targets for measuring shear directly.  
Note that 
without measuring the velocity field, it is impossible to 
be sure that a galaxy is rotating and therefore accurately  
elliptical rather than just having a moderate intrinsic   
aspect ratio. 

Figure\,\ref{fig4} shows the results of fits to a common shear field  
$\vec \gamma$
for simulated position and velocity data representing two disks with 
P.A.'s $45^\circ$ apart. The 
P.A.'s are fixed, and assumed to be determined accurately from the 
velocity data.  
The maximum probability of the fit is shown for each value of 
$\vec{\gamma}$ by solid lines, each of which is associated with 
a specific value of $i$ for the best fit that is 
shown by dashed lines for  
the galaxy with ${\rm P.A.} = -0.285$\,rad. 
There is a very significant degeneracy between the fitted values of 
$\vec{\gamma}$ and $i$. The direction of this degenerate strip in 
the $\vec{\gamma}$ 
plane is fixed by the P.A., and rotates by $2 \Delta$P.A.
if P.A. changes 
by $\Delta$P.A. Hence, observations of two galaxies that sample the 
same shear, with 
P.A.'s that differ by 
about 45$^\circ$ or 
135$^\circ$, can be used to determine the shear accurately 
and unambiguously, as shown by the overlapping solid 
contours in Figure\,4. The target galaxies should be at 
comparable 
redshifts, to ensure that they sample the same foreground structure 
and the nearer galaxy does not significantly lens the 
farther one. 

If more than two galaxies are 
observed, then the accuracy of the correspondence between the various 
tracks can be used as a test of the reliability of the method. 
Note that $\vec{\gamma}$ measurements from the 
rotation curves of several 
galaxies do 
not rely on a correlation technique. Hence  
spin or alignment correlations between pairs of galaxies 
(Crittenden et al.\ 2001; Lee \& Pen 2001) 
should not affect their accuracy,  
provided that the galaxy velocity fields  
are not distorted strongly by tidal interactions.  
Rather than determining a 
single average 
value of $\vec{\gamma}$ over each 10\,arcmin$^2$ area of sky, or
a correlation spectrum of $\vec{\gamma}$ on 
scales of 0'.5--10', 
it should be possible to determine a value of $\vec{\gamma}$ in a 
tube on the order of 
a few arcseconds wide 
along the line of sight to a 
specified point on the sky. 

Such a measurement would provide a direct test of systematic effects 
in shape-based statistical shear measurements by comparing their 
results at specific positions. The shear from 
individual mass concentrations, including subcritical groups of 
galaxies, could be 
mapped in detail without stacking groups 
(Hoekstra et al.\ 2001) by measuring the shear 
along closely spaced lines of sight around the target. 
It may also be 
possible to probe for significant structure in the shear field on 
arcsecond 
angular scales, as expected if dark matter halos 
contain a large amount of substructure, by measuring differences in the 
shear field determined for galaxies only a few arcseconds apart. 
\footnote{ 
Combining the measured CO rotation curve of highly inclined  
disks, which are often expected to strongly lens background galaxies  
(Blain, M\"oller \& Maller 1999), with the positions of  
serendipitous multiple background images 
is unlikely to provide more accurate constraints on cosmological 
parameters than observing 
multiple images in rich clusters 
(Golse, Kneib, \& Soucail 2002). Because 
both lensing and 
rotation-curve measurements are sensitive to the ratio of enclosed 
mass to radius, it is not possible to measure a 
`standard ruler' 
across the galaxy to determine $H_0$. 
}

\section{Requirements for detection}

In order to exploit this effect, it will be necessary to measure very 
accurate 
rotation curves of distant galaxies. 
The shear signal is significant only at moderate redshifts ($z > 0.3$; 
Bartelmann \& 
Schneider 2001), at which the total extents of galaxies are only a few 
arcseconds.  
Sub-0.1"
spatial resolution is thus necessary.
Measurements of 
velocity with 5\% 1-$\sigma$ accuracy should be achievable based on 
low-redshift CO and H{\sc i} observations of rotation curves traced by 
clouds with 10\,km\,s$^{-1}$ internal velocity dispersions and scatter from cloud to cloud. 
The errors expected from 100 simulated position-velocity 
data points of this accuracy are 
shown 
in Figure\,\ref{fig3}. 
Even in the most disturbed extra-nuclear 
molecular cloud 
regions, the maximum velocity dispersion is about 20\,km\,s$^{-1}$
(Bally et al.\ 1999). It is essential that the 
velocity field faithfully traces 
rotation, 
and is not due to outflowing gas moving at up to 
several 100\,km\,s$^{-1}$, as is possible for 
optical nebular emission lines. The required sensitivity and 
resolution could probably be achieved either by using AO IFU spectroscopy 
of near-IR 
absorption or recombination lines or by measuring high-/low-excitation 
CO lines using 
ALMA/SKA or H{\sc i} 
emission using SKA. 
We will consider the prospects for ALMA and SKA. 

The ALMA bands at 230 and 345\,GHz can be used to detect CO lines  
excited in star-forming regions of ordinary disk 
galaxies at $z \sim 0.5-1$. 
CO is likely to be distributed within the 
inner several kiloparsecs 
of the disk, probably more centrally concentrated than the 
stellar light; however, there is a possibility that the CO emission may be 
concentrated very 
close to the nucleus, in which case resolved ALMA observations would be 
impossible, and centimeter-wave 
SKA observations of CO(1-0) or H{\sc i} would be 
required.  
At 230\,GHz, CO(3-2)  
from $z \simeq 0.5$ and CO(4-3) from $z \simeq 1$ can be detected. 
At 345\,GHz, CO(5-4) can be detected from $z \simeq 0.65$. 
The expected luminosity can be calculated by scaling the 
high-resolution images of CO(1-0) from the low-redshift 
Berkeley--Illinois--Maryland Association Survey Of Nearby Galaxies (BIMA SONG; 
Regan et al. 2001) to higher redshifts and  
excitations or by scaling down the flux density of CO(3-2) detected in 
high-redshift ultraluminous galaxies (Frayer et al.\ 1999). The 
scaled-down high-redshift result implies a total integrated CO(3-2) 
line flux of 0.5\,Jy\,km\,s$^{-1}$ from $z=0.5$ for a Milky Way-like 
galaxy with 
a luminosity of about $6 \times 10^{10}\,L_\odot$. 
Scaling from CO(1-0) for NGC\,4414 from BIMA SONG, 
assuming the Large Velocity Gradient model presented in 
Blain et al.\ (2000), line fluxes of 0.43, 0.49 and 
0.16\,Jy\,km\,s$^{-1}$ are expected in 
CO(3-2) from $z=0.5$ at 230\,GHz, CO(5-4) from $z=0.65$ at 345\,GHz,  
and CO(4-3) from $z=1$ at 230\,GHz, respectively. 
ALMA has a 1$\sigma$ sensitivity of 6.1 and 11\,mJy\,km\,s$^{-1}$ in the
230 and 345\,GHz bands, 
respectively, for a 1\,minute observation:  
70$\sigma$, 44$\sigma$ and 27$\sigma$ detections are thus expected 
integrated over the whole disk 
of the galaxy in 1\,minute for these three lines, respectively. 
For the most promising case, CO(3-2) from $z=0.5$, this means that 
100 independent spectra on the galaxy could be measured 
at 10$\sigma$ significance in about $(100)^2 / 7^2$\,minutes or 
3.2\,hr. At 230\,GHz, ALMA has a maximum resolution of about 
30\,mas (on a 10\,km baseline), and therefore 
has sufficient 
spatial and spectral resolution to measure $\vec{\gamma}$. 
Note that the time required to make a coarse measurement of the 
total luminosity and 
extent of CO emission  
using ALMA is very 
much less than that required to produce a fully resolved image 
for shear measurements. Hence, only  
galaxies that seem to be regular rotating disks in short snapshot 
observations need be subjected to 
1--10\,hr integrations to measure 
accurate rotation curves. 
The morphology of the dust emission from the target is determined 
simultaneously, for comparison 
with that of radio emission and near-IR/optical 
starlight. 

The SKA detection limit 
for H{\sc i} emission from $z \simeq 1$ is several 10$^8\,M_\odot$ in many 
tens of hours of integration. 
Hence, it should be possible to detect 
several tens of 0".1 resolution elements across a high-redshift 
galaxy similar to the $5 \times 10^{9}\,M_\odot$ NGC\,4414 in a 
reasonable time at 1\,GHz, for  
baselines longer than about 600\,km. SKA has 
a much larger field of view than ALMA and   
will image a very large number of galaxies simultaneously. As discussed 
in its web-based science case (see footnote 1) SKA  
could carry out conventional 
`statistical' cosmic shear studies using 
just the  
shapes of galaxies. 

In order to estimate the signal-to-noise ratio for the velocity measurements, 
a reasonable figure of merit is the product of the resolved area 
of the galaxy on the sky (which determines the number of 
independent velocity measurements) and the size of the 
change in the projected line-of-sight velocity across the disk. 
These functions depend on cos\,$i$ and sin\,$i$ respectively.
Only in the most edge-on disks would CO emission be obscured by and 
blended with foreground clouds in the target galaxy, and therefore 
the total integrated flux in a CO line is not expected to depend on $i$. 
The figure of merit thus depends on sin\,$2i$ and is maximized at  
$i \simeq 45$\,deg: 
edge-on galaxies yield 
too few independent data points, while the velocity gradient in face-on 
galaxies 
is too small to be sure that the motion is rotational.  

The shear signal is expect to increase as $z^{\simeq 0.6}$ 
at source redshifts $z \simeq 0.5$--1 (Bartelmann \& 
Schneider 2001). Note that the 
integrated flux of a line decreases as 
$D_{\rm L}^{-2}$, and the number of resolved spectra decreases 
as $D_{\rm A}^{-2}$,  
where $D_{\rm L}$ and $D_{\rm A}$ are the luminosity and 
angular diameter distances, respectively.
$D_{\rm A} = D_{\rm L}(1+z)^{-2}$, 
leading to
$(1+z)^{-4}$ surface brightness dimming, and therefore it is easier to detect 
and resolve lines from galaxies at lower redshifts. A conservative 
figure of merit for the redshift dependence of the detectability of 
lines for measuring $\vec{\gamma}$ 
would be $D_{\rm L}^{-2}$, which decreases as 
$z^{\simeq -2.4}$ from $z \simeq 0.5$--1, 
much more steeply than $\vec{\gamma}(z)$ 
increases. This is before the likely systematic 
decrease in galaxy disk size/mass 
with increasing redshift is taken into account. 
Hence, in order to detect shear it is 
sensible to observe relatively low redshift galaxies: this 
favors 230\,GHz ALMA observations of CO(3-2) from $z \simeq 0.5$. 

\section{Caveats}

In order to detect the velocity structure of weakly lensed galaxies and 
determine $\vec{\gamma}$, observations with 
very high spatial and spectral resolution are required. The measurements 
need to be accurate to within about 5\%, and so low-level systematic
effects could potentially confuse and affect the results. 
We have illustrated the effect for a ring geometry.  
Spatially resolved spectroscopy of any disk built from 
a series of circular rings should allow this 
signal to be extracted. Even a galaxy with a 
series of irregularly distributed knots within a disk or spiral 
structure without radial inflow should allow an accurate rotation curve to 
be derived. If appreciable inflow does occur, 
then the galaxy would quickly be recognized as a poor target for
a deep rotation-curve measurement. Low-redshift 
galaxies observed in H{\sc i} can be well fitted by 
circular rotation; however, the P.A. can vary with radius (e.g.,  
Mulder \& 
van Driel 1993). This type of effect could prevent the 
detection of weak shear 
ring by ring within the galaxy. 

To measure shear, the detected velocities must be assumed to be 
associated with regular circular 
orbits. If the galaxy has a bar, warp, or 
$m_1$ asymmetry, 
then this model may not be adequate, and low-level corrections to 
a simple circular model may mimic weak lensing.  
Bars appear to be less common at high redshifts (Abraham et al.\ 1999); 
however, 
warps are likely to be more common since galaxy 
interactions are more common.  
At low redshifts, bars tend to affect the dynamics within 
2--3\,kpc of the centers of disks, while warps affect orbits outside 
10\,kpc. Their signatures should be easy 
to see after much shorter ALMA integrations than are required to 
detect accurate rotation curves. In order to assess the detailed 
feasibility of the proposed observations, 
very high resolution  
CO and H{\sc i} rotation curves at radii $\simeq 5$\,kpc 
at low and moderate redshifts 
will be required to check for systematic effects. 

\section{Conclusions}

The use of measured high-resolution rotation curves 
of distant disk galaxies to determine cosmic shear 
has been described. 
The effects 
of lensing on the velocity and position of gas clouds in the 
galaxy are different, and so adding velocity information can provide a much 
more accurate shear measurement than shape alone. 
By observing two or more galaxies at relative P.A.'s 
of about 45$^\circ$, a direct measurement of the 
shear can be made, without requiring correlations  
or averages of large numbers of faint galaxy images. 
Such observations are not possible at 
present, but will become so with AO IFU spectrographs, and future 
radio and millimeter-wave interferometers. Velocity measurements 
are essential in order to 
be sure that the observed galaxy is rotating (and has an exact 
intrinsic elliptical morphology) and to allow searches for 
systematic deviations from uniform rotation. 

\acknowledgements

I thank Eric Agol, Richard Ellis, 
Jean-Paul Kneib, Ole M\"oller, Priya Natarajan 
and Kartik Sheth 
for helpful conversations.


\begin{thebibliography}{}
\bibitem[Abraham et al.(1999)]{Abr} Abraham, R.\,G., Merrifield, M.\,R., 
Ellis, R.\,S., Tanvir, N.\,R., \& Brinchmann, J. 1999, \mnras, 308, 569
\bibitem[Bacon et al.(2000)]{BRE} Bacon, D.\,J., Refregier, A., \& 
Ellis, R.\,S. 2000, \mnras, 318, 625
\bibitem[Bally et al.(1999)]{Bally} Bally, J., Reipurth, B., Lada, C.\,J., 
\& Billawala, Y., 1999, AJ, 117, 410
\bibitem[Bartelmann \& Schneider (2001)]{BS} Bartelmann, M., \& 
Schneider, P. 2001, Physics Reports, 340, 291  
\bibitem[BFBS]{BFBS} Blain, A.\,W., Frayer, D.\,T, Bock, J.\,J., \& 
Scoville, N.\,Z., 2000, MNRAS, 313, 559. 
\bibitem[Blain, M\"oller, \& Maller (1999)]{BMM} Blain, A.\,W., 
M\"oller, O., \& Maller, A.\,H. 1999, \mnras, 303, 423
\bibitem[Crittenden et al.(2001)]{Critt} Crittenden, R.\,G., Natarajan, P., 
Pen, U.-L., 
\& Theuns, T., 2001, \apj, 559, 552
\bibitem[Fischer et al.(2000)]{Fisch} Fischer, P., et al. 2000, \aj, 
120, 1198
\bibitem[Frayer et al.(1999)]{Fray} Frayer, D.\,T., et al. 1999, \apj, 
514, L13
\bibitem[Golse, Kneib, and Soucail(2002)]{GKS} Golse, G., Kneib, J.-P., 
\& Soucail, G., 2002, \aap, in press (astro-ph/0103500) 
\bibitem[Hoekstra et al.(2001)]{Henk} Hoekstra, H., et al. 2001, \apj, 548, L5 
\bibitem[Lee, and Pen (2001)]{LP} Lee, J., \& Pen, U.-L. 2001, \apj, 555, L106
\bibitem[McKay et al.(2002)]{McKay} McKay, T.\,A., et al. 2002, \apj, submitted
\bibitem[Maoli et al.(2001)]{Maoli} Maoli, R., van Waerbeke, L., 
Mellier, Y., Schneider, P., Jain, B., Bernardeau, F., Erben, T., \& 
Fort, B. 2001, \aap, 368, 766 
\bibitem[Mulder, and Driel(1993)]{MD} Mulder, P.\,S., \& van Driel, W., 1993, 
\aap, 272, 63  
\bibitem[Narasimha, and Chitre(1993)]{NC} Narasimha, D., \& Chitre, S.\,M., 
1993, \aap, 280, 57
\bibitem[Pirzkal et al.(2001)]{Pirz} Pirzkal, N., et al. 2001, \aap, 
375, 351 
\bibitem[Regan et al.(2001)]{Reg} Regan, M.\,W., Thornley, M.\,D., Helfer, 
T.\,T., Sheth, K., Wong, T., Vodel, S.\,N., Blitz, L., \& Bock, D.\,C.-J.
2001, \apj, 561, 218 
\bibitem[Schneider et al.(1992)]{SEF} Schneider, P., Ehlers, J., \& 
Falco, E.\,E., 1992, Gravitational Lenses. Springer: Berlin 
\bibitem[van Waerbeke et al.(2001)]{vW01} van Waerbeke, L., et al. 2001, 
\aap, 374, 757
\bibitem[Wilson et al.(2001)]{Wilson} Wilson, G., Kaiser, N., \& Luppino, 
G.\,A., 2001, \apj, 556, 601 
\bibitem[Wittman et al.(2000)]{Witt} Wittman, D.\,M., Tyson, J.\,A., 
Kirkman, D., Dell'Antonio, I., \& Bernstein, G. 2000, Nature, 405, 143
\bibitem[Wootten]{Woo} Wootten, A. ed 2001, Science with the Atacama Large
Millimeter Array, PASP Conf. 
Ser. vol. 235
\end{thebibliography}
\end{document}